\documentclass[a4paper]{jpconf}

\usepackage{graphicx}
\usepackage{xspace}

\usepackage{aasjournals}

\usepackage{url}
\newcommand{\C}{$^{12}$C\xspace}
\newcommand{\Ox}{$^{16}$O\xspace}
\newcommand{\Nee}{$^{20}$Ne\xspace}
\newcommand{\Ne}{$^{22}$Ne\xspace}
\newcommand{\Ni}{$^{56}$Ni\xspace}

\newcommand{\pv}{\ensuremath{\phi}}

\newcommand{\msun}{$M_\odot$\xspace}

\newcommand{\unitstyle}[1]{\ensuremath{\mathrm{#1}}}

\newcommand{\second}{\unitstyle{s}}

\newcommand{\angstrom}{\mbox{\AA}} %Angstrom
\newcommand{\rhoDDT}{\ensuremath{\rho_{\rm DDT}}}

\begin{document}

\title{Cosmic Chandlery with Thermonuclear Supernovae}

\author{A~C~Calder$^{1,2}$, B~K~Krueger$^3$,
A~P~Jackson$^1$, D~E~Willcox$^{1}$, B~J~Miles$^4$ and D~M~Townsley$^4$}

\address{$^1$ Department of Physics and Astronomy, 
Stony Brook University, Stony Brook, NY 11794-3800, USA}
\address{$^2$ Institute for Advanced Computational Science,
Stony Brook University, Stony Brook, NY 11794-5250, USA}
\address{$^3$ XCP-2, 
Los Alamos National Laboratory, Los Alamos, NM 87545, USA}
\address{$^4$ Department of Physics and Astronomy, University of Alabama, Tuscaloosa, AL 35487-0324, USA}

\ead{alan.calder@stonybrook.edu}

\begin{abstract}
Thermonuclear (Type Ia) supernovae are bright stellar explosions, the light curves 
of which can be calibrated to allow for use as ``standard candles" for measuring 
cosmological distances. Contemporary research investigates how the brightness of an 
event may be influenced by properties of the progenitor system that follow from 
properties of the host galaxy such as composition and age.  The goals
are to better understand systematic effects and to assess the intrinsic
scatter in the brightness, thereby reducing uncertainties in cosmological
studies. We present the results from ensembles of simulations in
the single-degenerate paradigm addressing the influence of age and
metallicity on the brightness of an event and compare our results to
observed variations of brightness that correlate with properties of the
host galaxy. We also present results from ``hybrid" progenitor models
that incorporate recent advances in stellar evolution.
\end{abstract}

\section{Introduction}
Supernovae are bright stellar explosions that signal the violent deaths of stars. 
These events synthesize heavy elements, may be a site of r-process nucleosynthesis,
and create neutron stars and black holes, which are the building blocks of other 
interesting astrophysical systems such as x-ray binaries. While these events
may be divided into multiple classifications observationally, the impetus for
the majority of events is understood to be either the release of gravitational binding
energy from the collapsing core of a massive star or the release of nuclear
binding energy when a thermonuclear runaway incinerates one or more compact stars.
Accordingly, these two mechanisms are referred to as ``core collapse" and ``thermonuclear," 
respectively, and together these produce and disseminate the majority of heavy
elements found in the galaxy and thereby drive galactic chemical evolution.
Core collapse supernovae create neutron stars and black holes and may be
a site of r-process nucleosynthesis, while thermonuclear supernovae completely
disrupt the progenitor, leaving only an unbound remnant.
Thermonuclear supernovae are particularly useful
as distance indicators for cosmological studies, making understanding these
events and properties such as their intrinsic scatter important to fields
well beyond stellar astrophysics. This manuscript reports on studies
of systematic effects on the brightness of thermonuclear supernovae that follow
from the evolutionary history of the progenitor star and host galaxy,
with the goal of better constraining uncertainty in the intrinsic scatter
of these events.

\subsection{Thermonuclear Supernovae}
Thermonuclear supernovae are understood to follow from a thermonuclear
runaway involving roughly one solar mass of stellar material under degenerate
conditions. Discerning the setting(s) of these events, however, is
proving to be difficult. Thermonuclear supernovae are unique among
supernovae in that they are also very important for cosmological
studies because their observational properties (e.g.\ the light curve) allow
calibration of events for use as ``standard candles," i.e.\ objects of
known brightness that may be used as distance indicators. Using
thermonuclear supernovae in this capacity resulted in the discovery of
the accelerating expansion of the Universe and thus the mysterious
dark energy driving the
acceleration~\cite{riess.filippenko.ea:observational,
  perlmutter.aldering.ea:measurements,leibundgut2001}.

Observationally, thermonuclear supernovae are classified as ``Type Ia" following
the classification scheme of Minkowski \cite{minkowski41}. The classification is
based on observing strong Si lines and a lack of H in the spectrum
of an event. Theoretically, these events are understood to follow from
the thermonuclear incineration of $\simeq 1$ \msun of a mixture principally
composed of C and O under degenerate conditions. The burning converts much
of the material to iron-group elements (IGEs) including $\sim 0.6$ \msun
of radioactive \Ni, the decay of which  powers the light curve 
\cite{pankey62,colgatemckee69,colgatepetschekkriese80,kuchneretal94}. 
The ability to standardize the light curves is understood to follow 
from the fact that the source powering the light curve, radioactive \Ni, is also the principal
source of opacity \cite{Pinto2001The-type-Ia-sup}. The result is that the B-band magnitude of
the light curves of brighter events decline more slowly from peak
than those of dimmer events. This ``brighter is broader" result
is known as the Phillips relation~\cite{phillips:absolute}, and it allows the calibration
of light curves via a single-parameter ``stretch" function~\cite{howelletal+09}.

Largely motivated by cosmological studies, many contemporary campaigns
are observing Type Ia supernovae. Observations 
suggest that there may be systematic effects on the peak magnitude
of the light curve even after calibration \cite{Kirshner09}, which 
is a significant source of uncertainty in cosmological studies utilizing these
as distance indicators.
Spectral diversity raises the possibility of two populations, 
which may follow from formation channel or from properties of the host galaxy 
and/or the local environment within the host galaxy. 
Arguments are made for two populations~\cite{MannucciEtAl06,
RaskinEtAl09, wangetal2013} or for correlation principally with age
measured as the delay time from early star formation in a 
galaxy~\cite{gallagheretal+08,howelletal+09,neilletal+09,BrandtEtAl10}.

One key measure in these studies is the amount of metals, elements 
heavier than He, that were synthesized by previous generations of stars. 
The presence of metals influences the evolution of stars 
and stellar explosions by increasing the number of pathways for 
nuclear burning~\cite{Chamulak2006Laminar-Flame-A}.  The metallicity 
of a particular galaxy increases with age, and within a particular 
galaxy metallicity changes with location, generally increasing 
toward the densely populated interior. In addition, 
observations of specific events also can address the issue of progenitors,
with some bright observations suggesting more radioactive \Ni than
could be produced by degenerate C and O in a single star~\cite{howell+06,
scalzo+10,yuan+10,tanaka+10}.

Although many of these events have been observed and these events serve
as the premier cosmological distance indicators, 
a robust theoretical understanding remains elusive. It is widely 
accepted that compact stars known as white dwarfs (WDs) composed 
principally of C and O provide the nuclear fuel, but the progenitor
systems have not been conclusively identified and questions about
the mechanism of the explosion remain. 
The three most widely accepted proposed progenitor systems are:
a single near-Chandrasekhar-mass WD that has gained mass from
a companion (the ``single degenerate" scenario), the merger of two 
WDs, and the sub-Chandrasekhar-mass double-detonation scenario in 
which a detonation in an accreted layer triggers a detonation in the core 
of the WD \cite[and references therein]{calderetal2013}. For studies
of systematic effects, we have adopted the single-degenerate model.

\subsection{The Single-Degenerate Explosion Paradigm}

The single-degenerate scenario encompasses several explosion mechanisms invoking both 
deflagrations (subsonic burning fronts) and detonations (supersonic burning fronts).
The key to a successful explosion, the production of an unbound remnant with a stratified
composition structure similar to observed remnants \cite{mazzalietal2008}, is the density 
at which the material of the WD burns. Some of the earliest work addressing the 
single-degenerate progenitor explored pure detonations and found that the
result is the vast majority of the WD material burns to IGEs
because of the relatively high density of the WD~\cite{arnett.truran.ea:nucleosynthesis}. 
In this case, there is no composition stratification in the remnant, and,
accordingly, pure detonations have been ruled out as a viable explosion mechanism 
for some time. 

Pure deflagrations have also been explored as the explosion mechanism. In 
this case, the star reacts to the subsonic burning and expands, which lowers the density 
of the material prior to it being consumed. The density in the outer regions of the
star gets low enough to quench the burning, and the result is an incompletely-burned,
bound object that does not release enough energy to explain the majority of
observed Type Ia supernovae \cite{roepkeetal07}. Pure deflagrations may explain some unusual
events, however \cite{kromeretal2015}. 

Because the two extremes of pure deflagrations or pure detonations cannot 
robustly reproduce events with characteristics consistent with the observations,
researchers have investigated models evoking a combination of the two. These models
may be broadly described as delayed detonation, and a variety of scenarios have
been proposed. The classic delayed detonation model was introduced by 
Khokhlov \cite{Khokhlov1991Delayed-detonat,hoflich.khokhlov.ea:delayed,GameKhokOran05}, 
and variations include pulsational 
detonations \cite{ivanovaetal1974,Khokhlov1991Delayed-detonat,
arnettlivne94a, arnettlivne94b, bravogarcia-senz2006},
gravitationally confined detonation \cite{PlewCaldLamb04,Jordan2008Three-Dimension,
jordanetal2012}, and deflagration-to-detonation 
transitions (DDTs) \cite{1986SvAL,Khokhlov1991Delayed-detonat,NiemWoos97,Niem99,belletal2004,
fishjump2015}. We adopt a variation of the latter case, a deflagration-to-detonation transition
that occurs at the top of a rising plume of hot, burned material when it
reaches a threshold density. 

The physics by which a DDT may occur are not completely understood, but
ideas have been proposed. One idea is that at low densities the flame
enters a regime of distributed burning in which it is no longer
well-defined but the net burning rate is high enough that it is
effectively supersonic \cite{NiemWoos97}. Similarly, others argue
that when the flame reaches a certain fractal dimension, it is
effectively supersonic \cite{woosley90}. The Zel'dovich mechanism
posits that composition and temperature gradients may ``prepare" the
fuel in just the right way that it will detonate
\cite{zeldovichetal1970,KhokOranWhee97,jacketal2014}. Our approach
is to assume that when the top of a rising, Rayleigh-Taylor unstable
plume reaches a particular density, conditions are favorable and the 
transition occurs \cite{townsley.calder.ea:flame}.

\section{Simulation Instrument}

The simulations of thermonuclear supernovae we describe here were performed 
with a modified version of the Flash code, developed 
at the University of Chicago.\footnote{The source code used for
these studies is available as a package compatible with the
current Flash code from \url{http://astronomy.ua.edu/townsley/code.} } 
Flash is a parallel, adaptive mesh, multi-physics simulation code
developed first for nuclear astrophysics applications and subsequently
for high-energy-density applications~\cite{Fryxetal00,calder.curtis.ea:high-performance,
calder.fryxell.ea:on,flash_pragmatic,flash_evolution}. The modifications to the 
Flash code for supernova simulations comprise routines to describe
thermonuclear burning during both the deflagration and detonation
phases, as well as routines to describe the evolution of
the dynamic ash.

The significant disparity between radius
 of a white dwarf ($\sim 10^9\ensuremath{\;}{\ensuremath{\mathrm{cm}}}$) and 
the width of laminar nuclear flame at high densities 
($< 1\ensuremath{\;}{\ensuremath{\mathrm{cm}}}$)
requires the use of a model during the deflagration phase because
even with many orders of refinement on an adaptive mesh, a
macroscopic simulation of the event cannot resolve the 
actual flame front.  Our model propagates an artificially
broadened flame with an advection-diffusion-reaction (ADR)
scheme~\cite{Khok95,VladWeirRyzh06} via evolution of 
a reaction progress variable $\phi$, where $\phi=0$ indicates unburned fuel
and $\phi=1$ indicates burned ash. The advection-diffusion-reaction equation
is 
\begin{equation}
  \label{eq:ard}
  \partial_t \pv + \vec{u}\cdot\nabla \pv = \kappa \nabla^2 \pv + 
\frac{1}{\tau} R\left(\phi\right) ,
\end{equation}
where $\vec{u}$ is the velocity of the fluid, $\kappa$ is the
diffusion coefficient, $\tau$ is the reaction timescale, and $R(\phi)$ is
a non-dimensional function. The diffusion and reaction parameters
$\kappa$, $\tau$, and $R(\phi)$ are tuned to propagate the reaction
front at a prescribed speed.  We use a modified version of the KPP
reaction rate discussed by~\cite{VladWeirRyzh06}, the ``sharpened KPP
reaction,'' with $R\propto(\phi-\epsilon)(1-\phi+\epsilon)$, where
$\epsilon \simeq 10^{-3}$.  This scheme has been shown to be
acoustically quiet, stable, and to give a unique flame speed
\cite{townsley.calder.ea:flame}. For the input flame speed, we use
tabulated flame speeds from direct numerical simulations of
thermonuclear burning~\cite{timmes92,Chamulak2007The-Laminar-Fla} and
boost these to account for enhancements to the burning from unresolved
buoyancy and background
turbulence~\cite{Khok95,gamezo.khokhlov.ea:thermonuclear,townsley.calder.ea:flame,jacketal2014}.

The ADR scheme describes a model flame, but more is needed to adequately 
describe the
burning in the white dwarf.  The burning of C and O under degenerate conditions 
can be described by three stages. First, C is consumed producing
reaction products close to Mg. Then O is consumed along with the ashes of C burning,
which produces a mixture of silicon group and light elements that is in a
statistical quasi-equilibrium~\cite{ifk1981,khok1981,khok1983}. Finally 
the silicon-group nuclei are converted to IGEs, reaching full nuclear statistical 
equilibrium (NSE).
Each of these stages is described with separate progress variables and 
separate relaxation times derived from full nuclear network calculations
\cite{Caldetal07,townetal2016}. 

In both the quasi-equilibrium and full statistical equilibrium, the creation of light elements
via photodisintegration balances the creation of heavy elements via fusion, thus maintaining the 
equilibrium. Energy can continue to be released as the
relative abundances change due to hydrodynamic motion, primarily occurring in rising 
plumes of hot burned material, which changes the local conditions such as density
and temperature and thereby adjusts the equilibrium. Electron capture also influences
the evolution in three ways: by shifting the binding energy of the material and thereby changing the
temperature due to released energy, by changing the Fermi energy and thus the pressure, and
by emitting neutrinos that escape and remove energy. Also, electron captures neutronize the
material, which produces more neutron rich iron-group material at the expense of \Ni. As
we will see below, this process has a significant impact on the \Ni yield and thus the
brightness of an event. Our flame capturing scheme incorporates these effects via
tabulated results from NSE calculations, the details of which
may be found in~\cite{SeitTownetal09}. With all of this included
physics, our scheme is able to describe dynamic evolution of the ash in addition
to the stages of C-O burning.

In addition to modeling the burning during the deflagration phase, our burning
scheme also describes detonations~\cite{Meaketal09,townetal2009} via progress
variables. In this case, we use thermally-activated burning with the actual
temperature-dependent rate for C consumption to allow a propagating shock to 
trigger burning, i.e.\ to propagate a detonation front. In our scheme, the
propagating detonation is able to describe the same stages of C burning as
the deflagration case, including the relaxation into 
NSE~\cite[and references therein]{townetal2016}.

\section{Systematic Effects on the Brightness of an Event}

\begin{figure*}[!ht]
  \begin{minipage}{0.32\textwidth}
    \includegraphics[width=\linewidth]{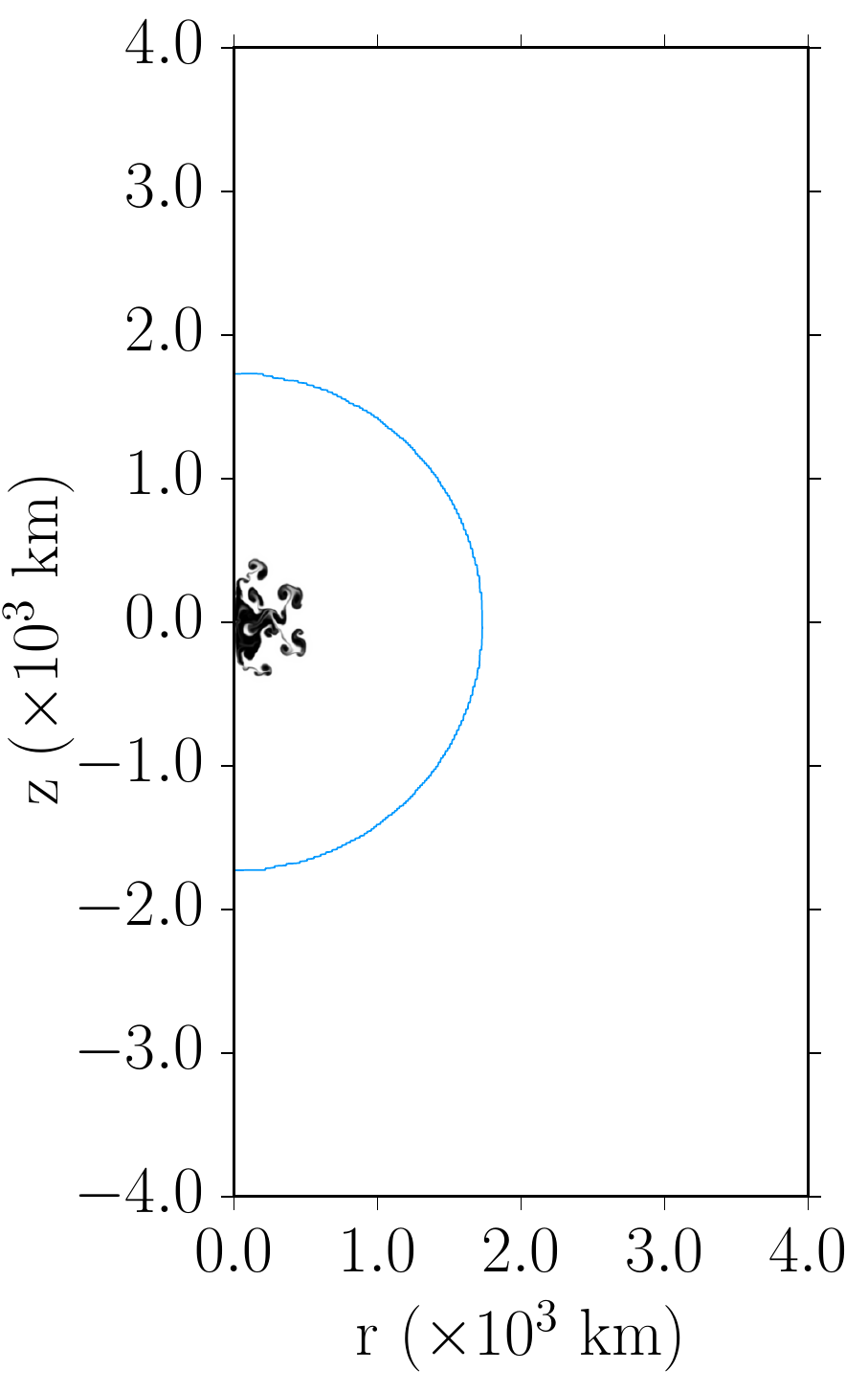}
  \end{minipage} \hfill
  \begin{minipage}{0.32\textwidth}
    \includegraphics[width=\linewidth]{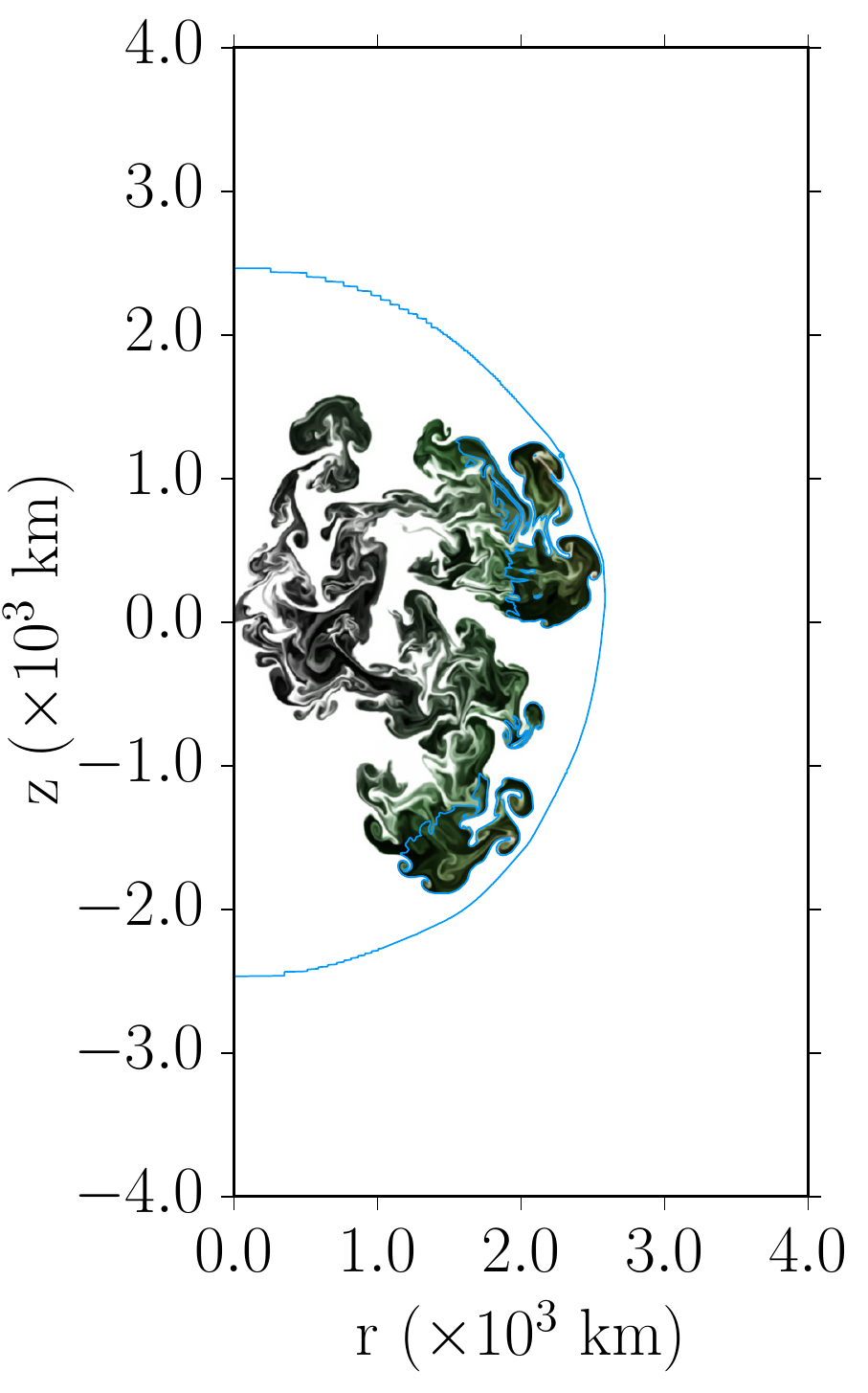}
  \end{minipage} \hfill
  \begin{minipage}{0.32\textwidth}
    \includegraphics[width=\linewidth]{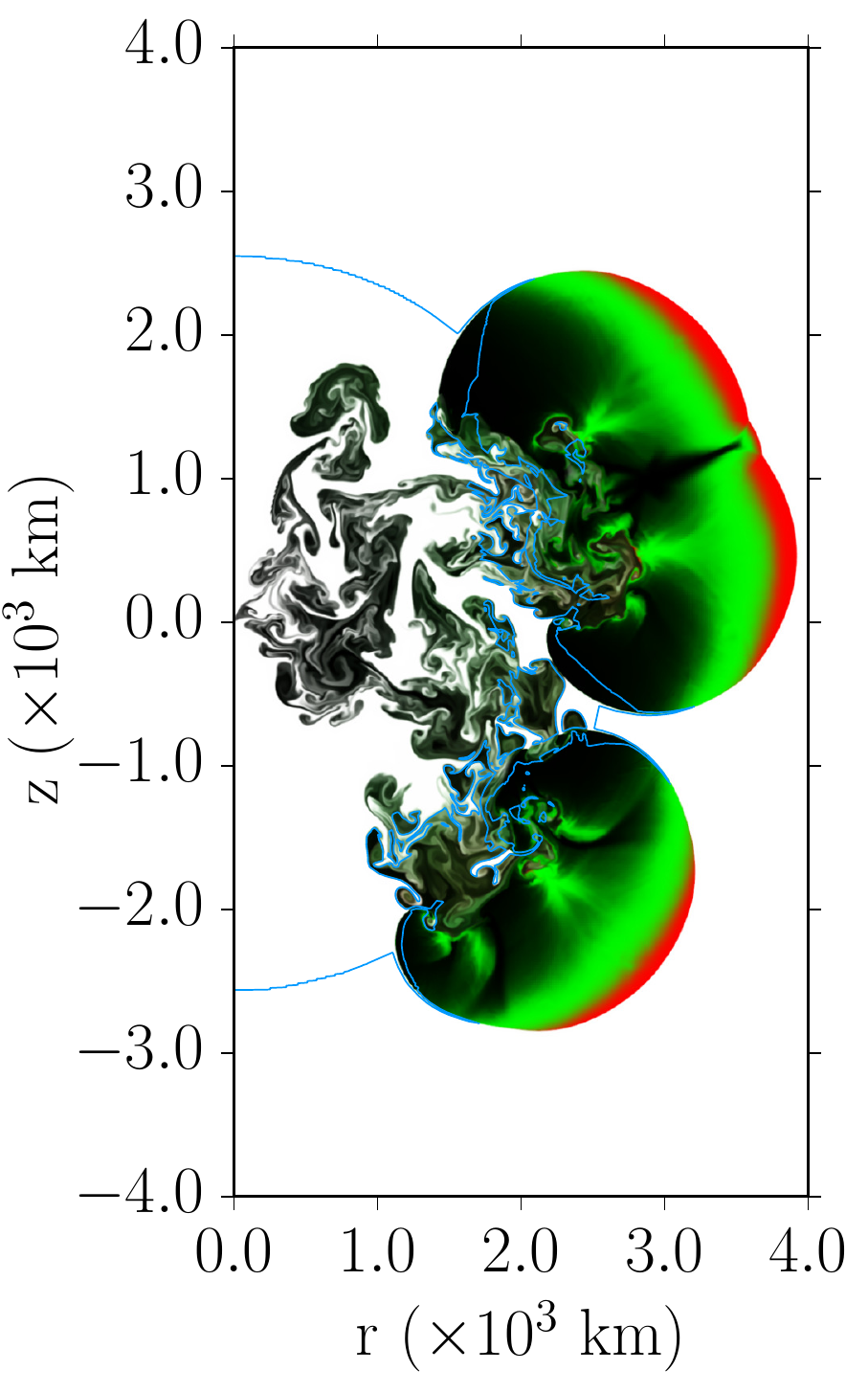}
  \end{minipage} \caption{\label{fig:co_ddt} Deflagration to detonation burning progress shown 
    for a C-O WD with $\mathrm{\rhoDDT=10^{7.2}~g/cm^3}$ (blue
    contour).  Unburned \C \& \Ox fuel is shown in white, and ash from
    \C-burning is shown in red. Matter in a quasi-nuclear statistical
    equilibrium state (primarily intermediate-mass silicon-group
    elements) is represented in green. Finally, matter in nuclear
    statistical equilibrium (IGEs and $\alpha$-particles) is shown in
    black.  From left to right, the burning is shown at 0.75~\second,
    1.74~\second, and 1.84~\second\ after the deflagration is initiated.}
\end{figure*}

We investigate systematic effects on the brightness of thermonuclear 
supernovae in the single-degenerate, DDT paradigm (described above).
Figure \ref{fig:co_ddt} illustrates the progression of thermonuclear burning
during a simulation in the DDT paradigm with a massive C-O progenitor WD. The left panel shows the flame
during the early deflagration stage with the obvious development of fluid
instabilities. The middle panel shows the flame as the first rising bubble
reaches the DDT threshold density (marked by the contour). The third panel
shows the propagation of two detonations. In all panels, the color scheme
shows the composition.

We note that an important part of our methodology is to perform
suites of simulations in order to investigate statistically-significant trends.
In \cite{townetal2009} we presented a theoretical
framework for these investigations utilizing multiple realizations from
randomized initial conditions.  Our results have shown that the problem
of thermonuclear supernovae in the single-degenerate DDT paradigm is deeply
nonlinear, particularly during the deflagration phase, which involves competition
between the growth of fluid instabilities and evolution of the flame. Accordingly,
these models exhibit a significant degree of variability, necessitating a
statistical approach \cite{scidac}. Finally, we note that all of the supernova
explosions presented here are from two-dimensional simulations. We found the
large number of simulations needed in ensembles necessitated the use of
two-dimensional models to be computationally tractable.

\subsection{Effect of Composition} 

The simplest description of burning in a C-O WD is the
alpha chain, a series of alpha capture reactions of the form
$X\left(\alpha,\gamma\right)Y$ going through symmetric nuclei ($n = p$)
from \C and \Ox  all the way to \Ni.  The presence of metals has been known
for some time to boost burning rates in WDs by offering additional
burning channels beyond just these basic reactions. In addition, metals
are neutron rich, and the neutron excess tends to drive the burning
away from symmetric nuclei. The result is that the presence of metals
affects the brightness of an event in two ways- faster flame speeds
lead to different dynamics during the explosion and neutron excess
suppresses \Ni production. 

Metal enrichment of the progenitor WD follows from two sources. First,
metals present in the massive star that evolved to become the WD will
be present in the progenitor WD. Second, convective burning or
``simmering" as the progenitor WD approaches the Chandrasekhar limit
synthesizes metals. In simulations, we use \Ne as a proxy for neutron
rich metals in the WD progenitor originating from both sources.

\begin{figure*}[!ht]
  \begin{minipage}{0.48\textwidth}
    \includegraphics[width=\linewidth]{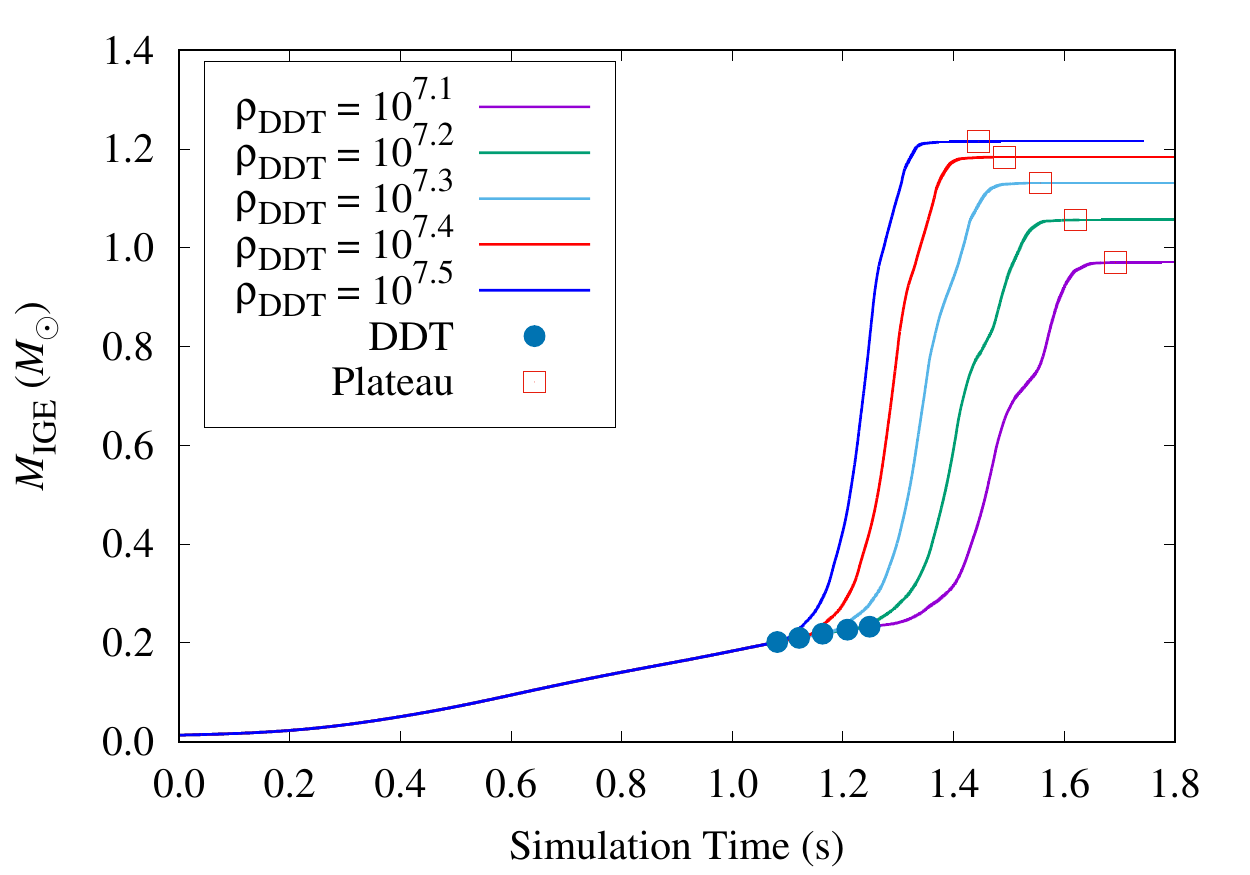}
  \end{minipage} \hfill
  \begin{minipage}{0.48\textwidth}
    \includegraphics[width=\linewidth]{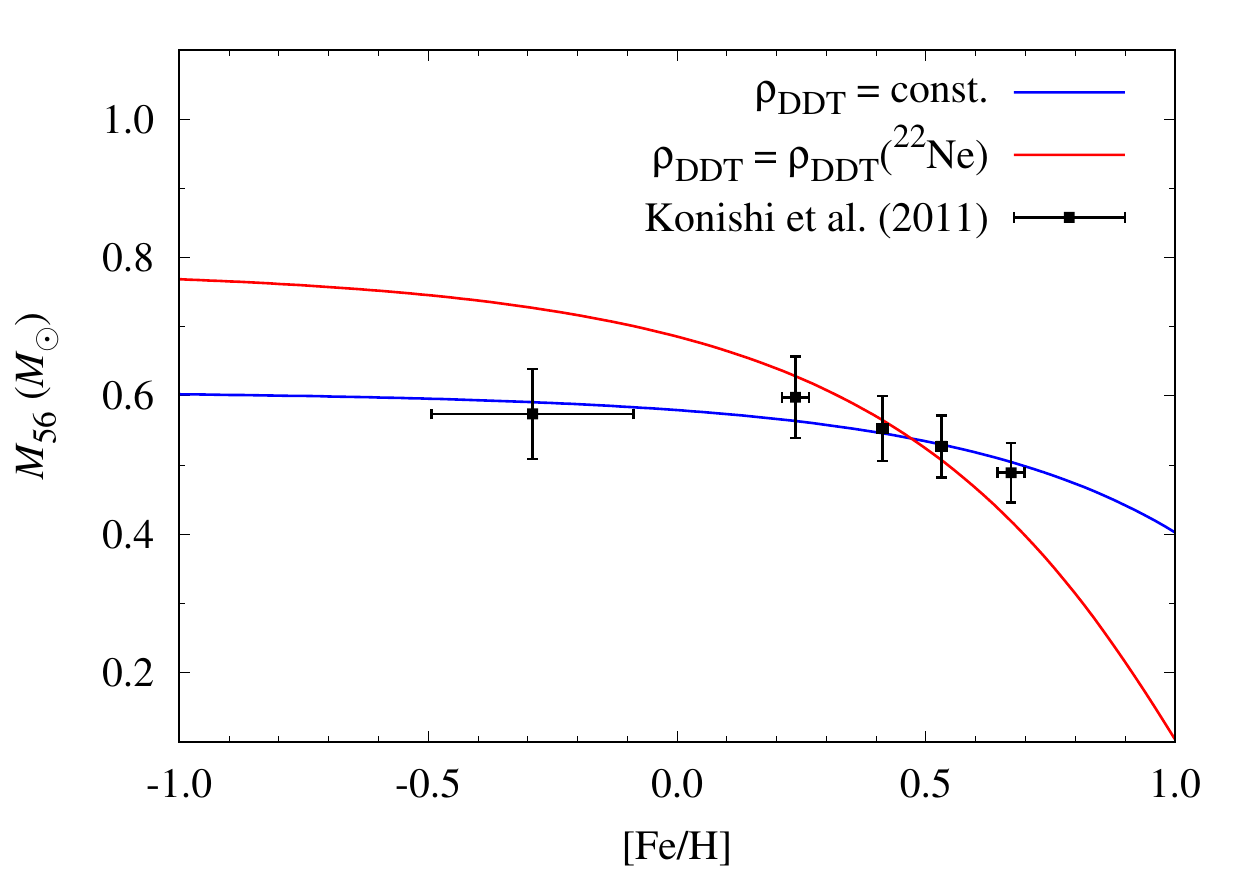}
  \end{minipage} \caption{\label{fig:metal} Left panel: Plot of IGE yield vs.\ time from simulation
of the same realization with the DDT density changing to account for
increased flame speeds due to increased metallicity.  Right panel:
Plot of produced mass of \Ni vs. metallicity comparing simulations to the
observational results of \cite{konishietal2011} (black points). Both
panels adapted from \cite{jacketal2010}.}
\end{figure*}

Based on essentially a counting argument,
\cite{timmes.brown.ea:variations} found that the presence of \Ne
directly modifies the \Ni abundance during thermonuclear burning of
a WD. Our models allow the investigation of the (nonlinear) effects
of metallicity on hydrodynamic models. As noted above, metallicity
boosts the burning rate by providing an increased number of reaction
channels. Accordingly, the flame speeds in explosion models are increased.
The increased flame speed changes the dynamics of the deflagration phase
and also will change the DDT density, which follows from a competition
between the development of fluid instabilities at the flame front and
fire polishing. In \cite{townetal2009}, we explored the role of \Ne, 
incorporating its effect on progenitor structure, laminar flame speed,
and energy release into simulations. We found that these influence the duration of the
deflagration phase, which in turn influences the amount of expansion prior
to the detonation that incinerates the star. The amount of expansion
determines the overall density structure, and in particular the amount
of mass at densities high enough to burn all the way to IGEs. Thus we
found metallicity does influence the yield of \Ni, but we found that
these effects were relatively weak. In \cite{jacketal2010}, we included
the effect of the changing flame speed on the DDT density. In this case,
we found a more drastic effect on the duration of the deflagration phase
and, accordingly, on the yield of \Ni.
Figure~\ref{fig:metal} shows results from these two studies along with observational results
for comparison.

\subsection{Effect of Central Density} 

In \cite{Krueger2010On-Variations-o} and subsequently in \cite{kruegetal12} we investigated the
role of central density on the yield of \Ni and thus the brightness. Our study constructed 
models with varying central densities and simulated explosions from these. We found that 
the overall production of IGEs is statistically independent of progenitor central density, but the
mass of stable IGEs is tightly correlated with central density, with a decrease in the production of
\Ni at higher central densities. Thus our results indicate that  
progenitors with higher central densities produce dimmer
events. Our understanding of the source of this trend is the increased neutronization that
occurs due to the faster rates of weak reactions at higher densities.  
By applying the relationship of Lesaffre, et al.\ \cite{Lesaffre2006The-C-flash-and}, 
which relates the duration
of WD cooling prior to the onset of accretion to the central density of the progenitor at
ignition, we were able to obtain a relationship between the central density and the cooling time,
a significant fraction of the age of the progenitor. With this relationship, we then had relationships
between the cooling time and the masses of IGEs and \Ni.

\begin{figure*}[!ht]
  \begin{minipage}{0.48\textwidth}
    \includegraphics[angle=-90,width=\linewidth]{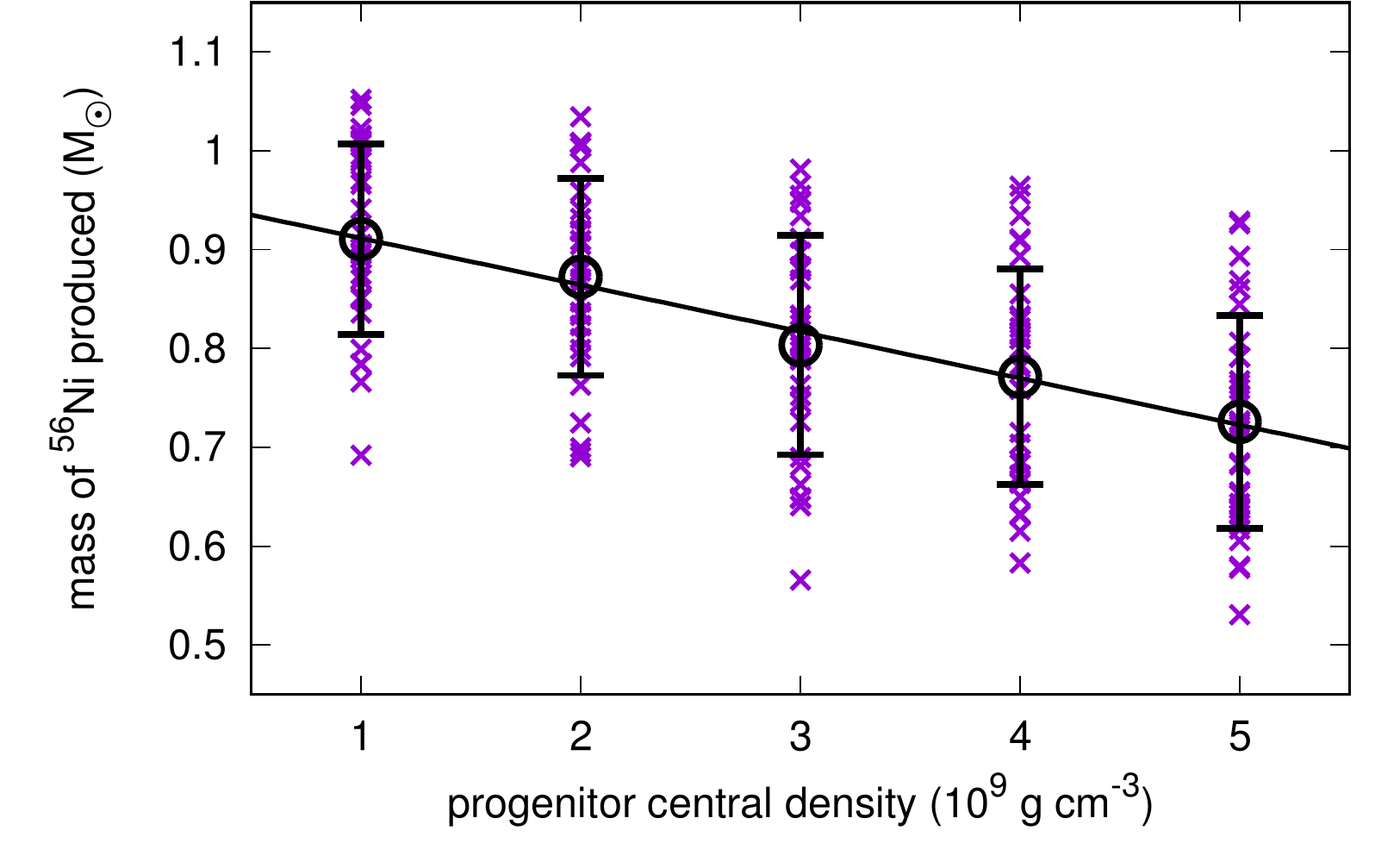}
  \end{minipage} \hfill
  \begin{minipage}{0.48\textwidth}
    \includegraphics[width=\linewidth]{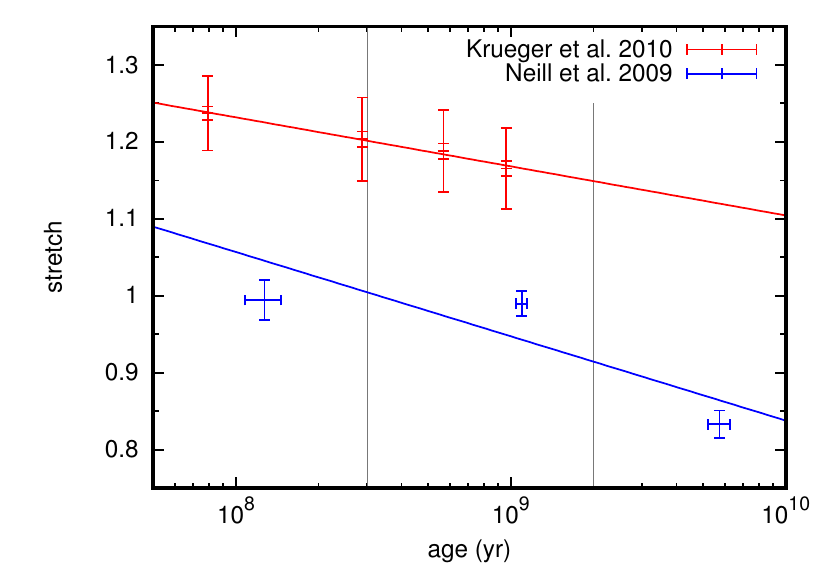}
  \end{minipage} \caption{\label{fig:neill} Left panel: Mass of \Ni as a function of
progenitor central density for 30 simulations at each central density. The black lines 
are the best-fit trend lines with averages and standard deviations marked by the
circles and vertical error bars. Right panel: Plots of stretch 
vs.\ age with 
the standard deviation, the standard error of the
mean, and a best-fit trend line. In blue are results from 
Figure 5 of \cite{neilletal+09} along with a best-fit trend line, and  
the vertical gray lines mark the cuts between bins. 
The overall offset to larger stretch in our simulations follows 
primarily from the choice
of DDT density. Adapted from \cite{Krueger2010On-Variations-o}. }
\end{figure*}

Results from the central density study are shown in Figure \ref{fig:neill}. The left panel shows 
the mass of \Ni produced in the explosion from progenitors at 5 central densities for 
30 realizations each. The right panel
shows relationship between age and stretch, a measure of the brightness when considering the
Phillips relation~\cite{howelletal+09}. Also shown are the observational results of Neill, 
et al.~\cite{neilletal+09} for comparison. An obvious result to note is that the trends agree, but
our stretches or yields are consistently higher. While evidence suggests that explosions
in the single-degenerate paradigm tend to be bright \cite{fishjump2015}, we attribute the
difference as primarily due to our choice of DDT density for the suite of simulations.

\subsection{Explosions from ``Hybrid" Progenitors} 

Recent work in stellar evolution posits the existence of ``hybrid"
C-O-Ne white dwarfs consisting of a C core surrounded by a mixture of
O and Ne. The structure of these white dwarfs follows from convective
boundary mixing quenching a C-flame before it can consume all of the C
in the late stages of a super asymptotic giant branch star's
life~\cite{denissenkovetal2013}. C burning produces \Nee that,
because of the convective boundary, remains at a larger radius than the
unburned C. The result is a hybrid C-O-Ne WD with more mass than a C-O
WD (and thus less mass is needed from accretion to approach the
Chandrasekhar limit)~\cite{denissenkovetal2015}.  During accretion, C
burning commences and drives convection that mixes and removes this
layering.  The result of this simmering is a progenitor WD with a C-depleted
core that is cooler than the surrounding material due to energy loss
from neutrino emissions. It differs from a C-O progenitor in its
temperature profile and because it has a significant amount of \Nee.  
Simulating thermonuclear supernovae from these hybrid progenitors thus 
required adapting the burning
module (described above) to account for the \Nee
\cite{willcoxetal2016}, and these adaptations were incorporated
into the source code to be distributed.
We note that studies of these hybrid models continue, and very recent
work \cite{brooksetal2016} suggests that mixing during the cooling phase of the WD before
the onset of accretion plays an important role and may produce progenitors
different in composition and structure from the hybrid models we employed in
our study. 

\begin{figure*}[!ht]
  \begin{minipage}{0.32\textwidth}
    \includegraphics[width=\linewidth]{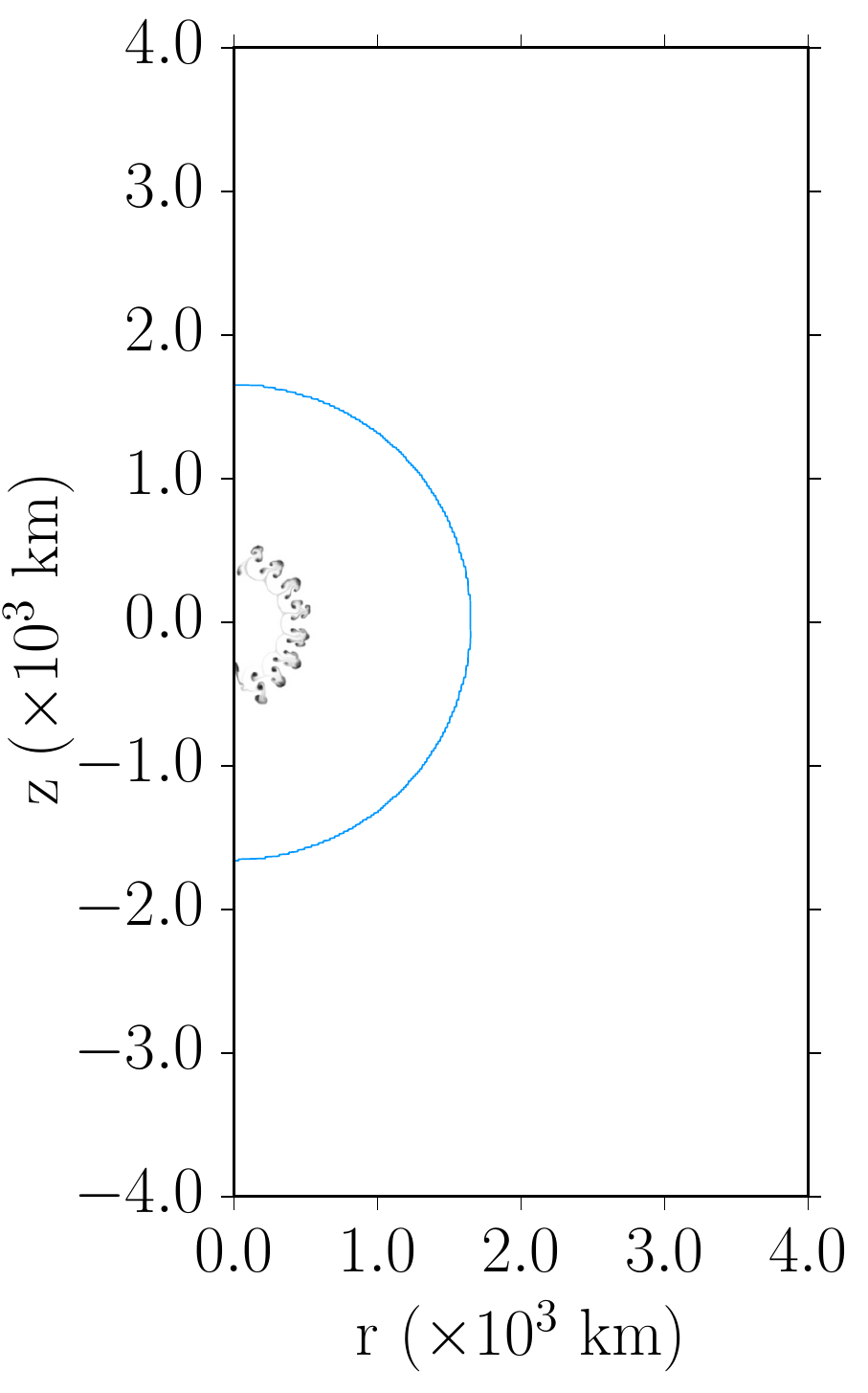}
  \end{minipage} \hfill
  \begin{minipage}{0.32\textwidth}
    \includegraphics[width=\linewidth]{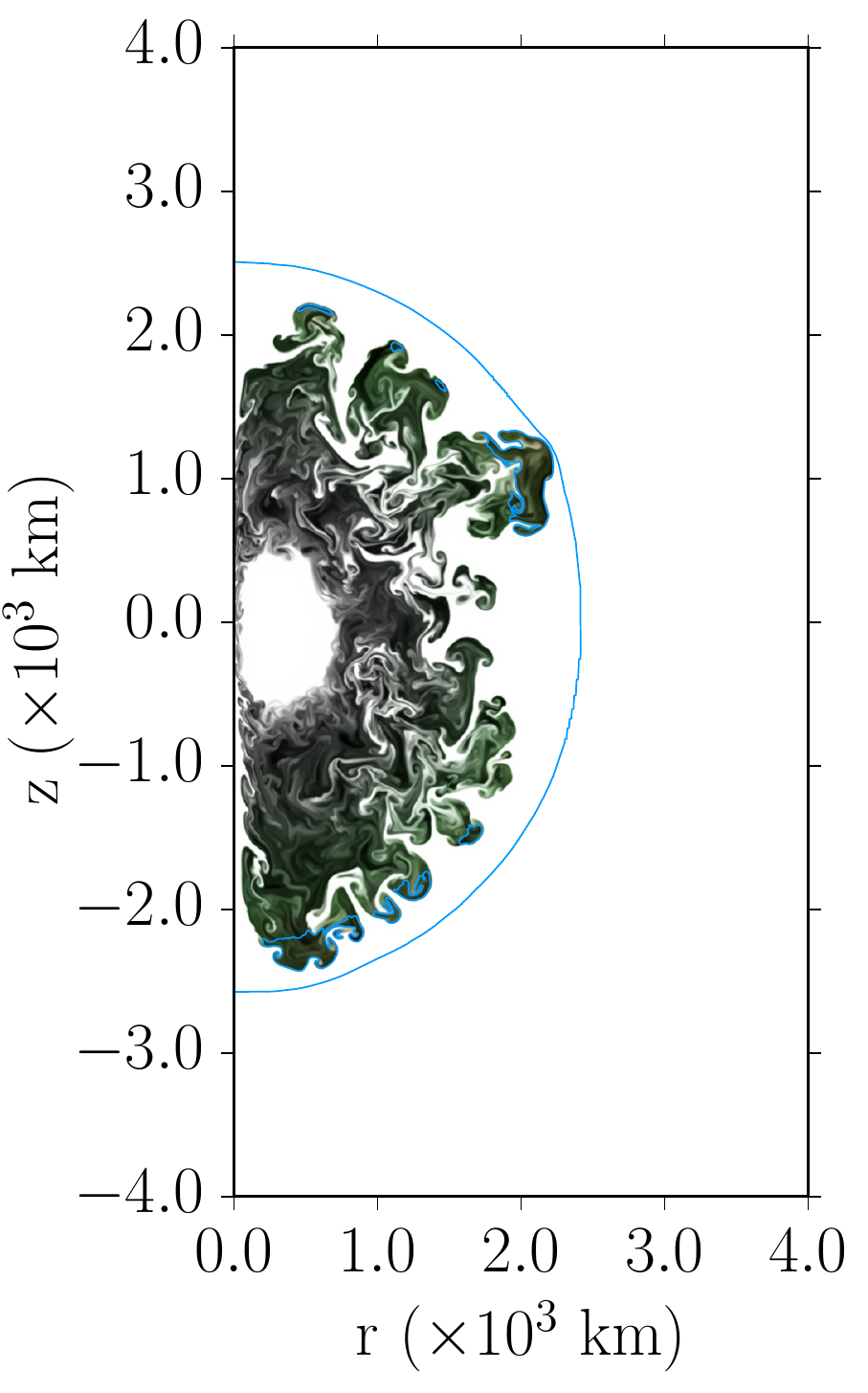}
  \end{minipage} \hfill
  \begin{minipage}{0.32\textwidth}
    \includegraphics[width=\linewidth]{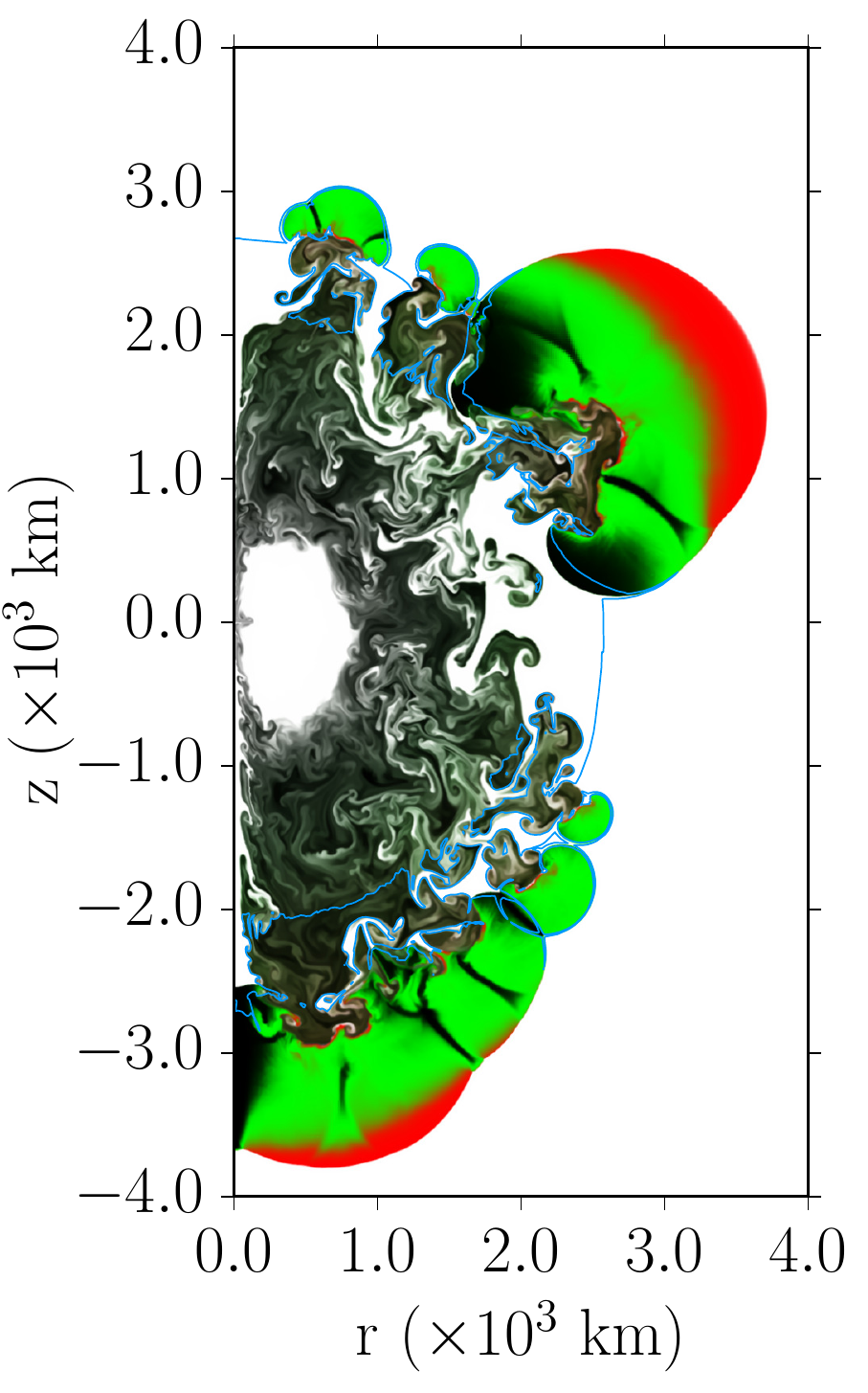}
  \end{minipage} \caption{\label{fig:cone_ddt} Deflagration to detonation burning progress shown 
for a C-O-Ne hybrid WD with $\mathrm{\rhoDDT=10^{7.2}~g/cm^3}$. The
color scheme is the same as that of Figure \ref{fig:co_ddt}.  From
left to right, the burning is shown at 0.50~\second, 1.48~\second, and
1.60~\second\ after the deflagration is initiated.}
\end{figure*}

We investigated the viability of these hybrid progenitors with a 
series of explosion simulations in \cite{willcoxetal2016}.
Figure \ref{fig:cone_ddt} shows the progression of thermonuclear
burning during a simulation similarly to the C-O
case illustrated in Figure \ref{fig:co_ddt}. The panels show the early
deflagration phase (left), the moment of ignition of the detonation
(center) and the propagation of the detonation (right). An obvious
difference between the hybrid progenitor and the C-O case is the cool,
C-depleted core, which shows little burning (the ``hole'' in the
center of the star seen in each panel) because the deflagration is
born in the high-temperature shell above the core. By the end of the
simulation, however, these models eventually consume the core during
the detonation phase. A noticeable difference from the C-O models is
this delay in burning the inner core.

\begin{figure*}[!ht]
  \begin{minipage}{0.48\textwidth}
    \includegraphics[width=\linewidth]{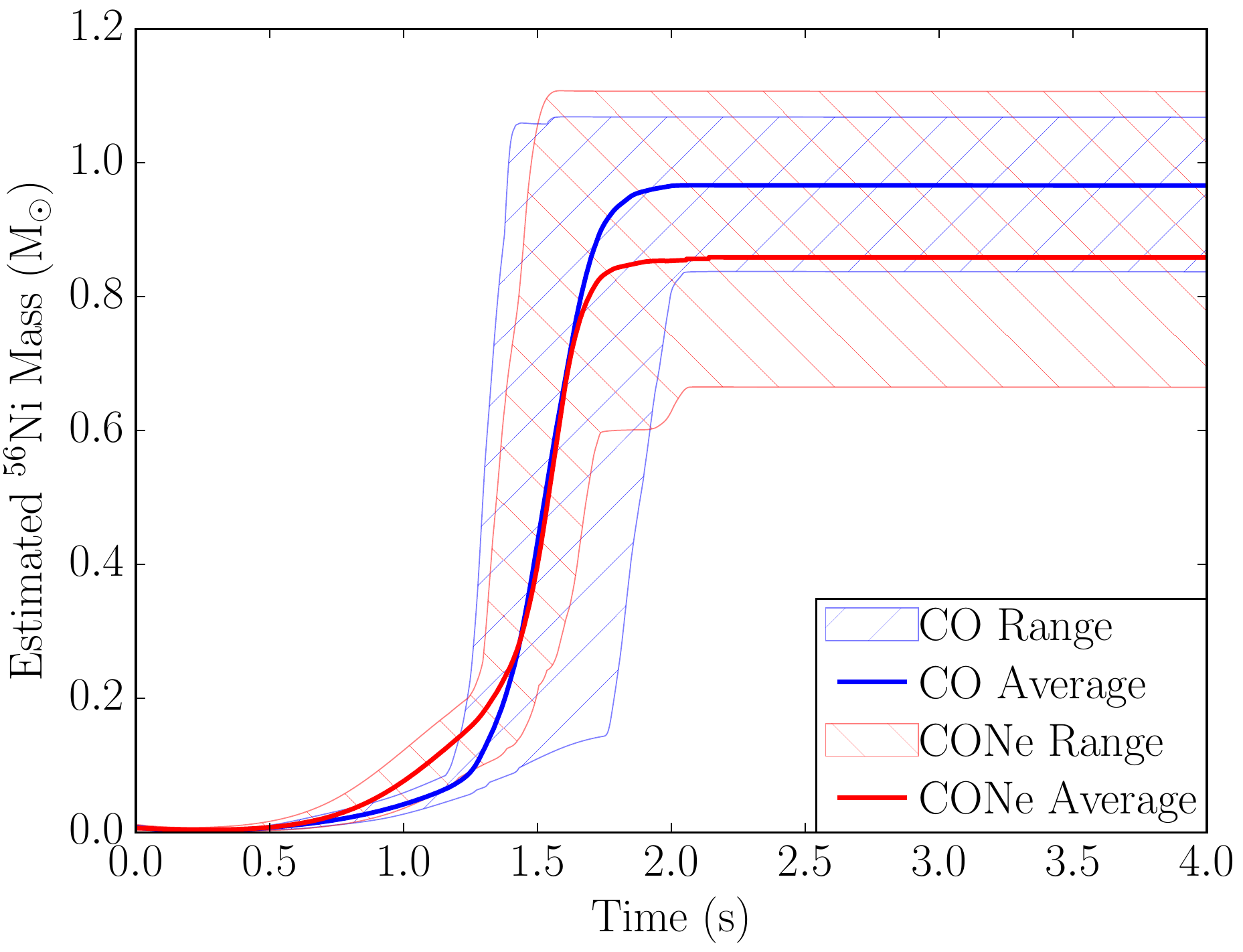}
  \end{minipage} \hfill
  \begin{minipage}{0.48\textwidth}
    \includegraphics[width=\linewidth]{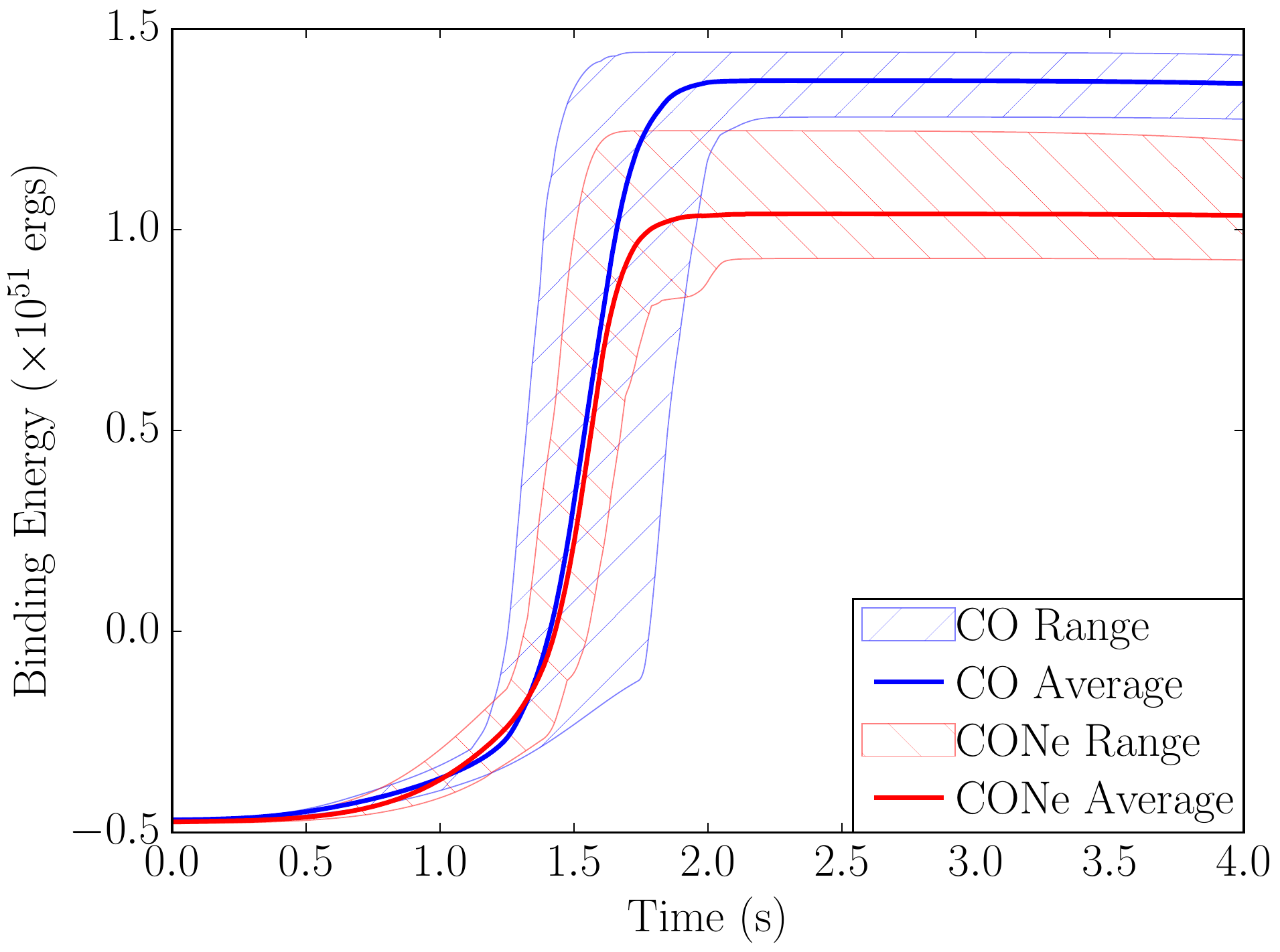}
  \end{minipage} \caption{\label{fig:cone_results} \Ni yield and gravitational binding energy
   from suites of simulations of explosions from hybrid C-O-Ne progenitors. The left panel shows
   the yield of \Ni from suites of simulations from C-O progenitors (very similar to those
   of \cite{kruegetal12}) and from C-O-Ne hybrid progenitors. 
   The right panel shows the binding energy from the two suites. In both panels, the 
   range of results are marked by the hashed regions. Adapted from \cite{willcoxetal2016}.}
\end{figure*}

We found that while there can be considerable variability in the outcome, there 
are statistically significant trends, with hybrid C-O-Ne models having a lower \Ni yield 
and ejecta with lower kinetic energy than similar C-O models. 
Figure \ref{fig:cone_results} illustrates these trends,
with the left panel presenting estimated \Ni yields and the right panel
presenting gravitational binding energy.
We concluded that these hybrid progenitors are viable
for supernova explosions.

\subsection{Analysis via Synthetic Spectra and Light Curves} 

The flame capturing and energetics scheme described above is sufficient to capture
the bulk energetics of the explosion and to estimate the yield of \Ni and IGEs. 
Making accurate calculations of the abundances synthesized in an explosion, however,
requires much higher fidelity calculations of the thermonuclear burning. We address
this issue by postprocessing fluid element histories (temperature and density) from 
Lagrangian tracer particles embedded in the hydrodynamic flow during a supernova 
simulation. In \cite{townetal2016} we presented a new method for this process with 
particular attention given to reconstructing the fluid element history within
the artificially-thickened model flame. We provided details of the method and
verification tests demonstrating that for the problem of deflagration and 
detonation fronts propagating in a uniform medium, we find good agreement 
between the post-processing method and a direct simulation of the burning. 

With these accurate, detailed abundances, we were able to generate
synthetic spectra from our post-processed explosion models to compare
with observations in order to explore how signatures of metallicity in
the spectral features may be used to constrain the metallicity of the
progenitor WD \cite{milesetal2016}. We performed explosion simulations
from progenitors for a range of metallicities, and from the spectra
were able to find correlations between the progenitor metallicity and
the strength of certain spectral features. Specifically, a Ti feature
near 4200 \angstrom\ and an Fe feature near 5200 \angstrom\ show a
trend of increasing width with increasing metallicity. These results
suggest the ability to observationally differentiate between
progenitor metallicities.  

\begin{figure}[!ht]
\centering
\includegraphics[angle=0, width=0.8\textwidth]{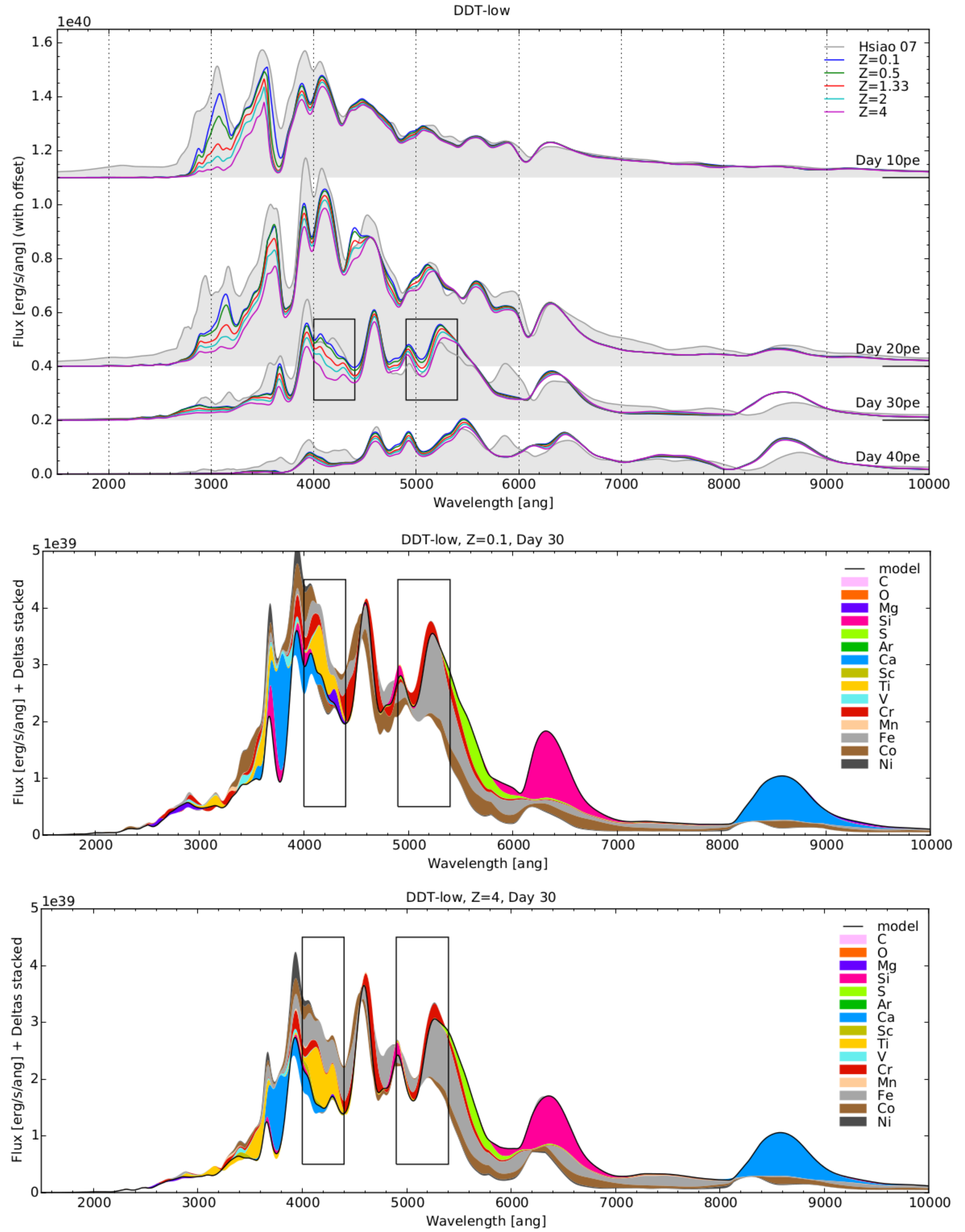}
\caption{\label{fig:knock} Spectra from a suite of simulations at
  varying metallicity at 10, 20, 30, and 40 days after the explosion
  illustrating evolution (top panel). Also shown are ``knockout"
  spectra from one DDT realization simulated with metallicity
  $\mathrm{Z/Z_{\odot} = 0.1}$ (middle panel) and
  $\mathrm{Z/Z_{\odot} = 4.0}$ (bottom panel) at 30 days after the
  explosion.  Contributions from line opacity of individual elements
  to the full spectrum (black solid lines) are represented by colored
  patches. These are calculated by removing the line opacities from a
  single element and recalculating the spectra while keeping the
  temperatures and population numbers fixed. Emission contributions
  are represented by colored patches below the full spectra, and
  absorption contributions are represented by patches above the full
  spectra. The spectral features highlighted with black boxes indicate
  two potential spectral indicators for progenitor
  metallicity. Adapted from \cite{milesetal2016}.  }
\end{figure}

Figure \ref{fig:knock} shows synthetic
spectra from post-processed remnants derived from simulations at
different metallicities. The top panel shows spectra at 10, 20, 30,
and 40 days after the explosion.  The middle 
and bottom panels show spectra from post-processed remnants 
at 30 days after the explosion that were derived from simulations
at low and high metallicities, respectively.  The spectra at
the two metallicities are ``knock out" spectra calculated with and
without the contribution to the opacity from an individual transition
line \cite{vanrossum2012}.  The two knock out spectra are from the same DDT
realization simulated at the two  metallicities.  Changes in the
differences between the model spectra and ``knock out'' spectra serve
as a diagnostic for changes following from varying metallicity.

\section{Conclusions}

We have developed a detailed flame-capturing and energetics scheme to
model thermonuclear supernovae and coupled it to post-processing techniques
to calculate detailed abundances and synthetic spectra and light curves. 
With these tools, we have investigated systematic effects on the brightness 
of thermonuclear supernovae (in the single-degenerate DDT paradigm)
that are thought to follow from the evolutionary history of the
progenitor star and host galaxy.  We investigated the effects of age
and composition and our results are in rough agreement with observed
trends. We also were able to quantify variations within the suites
of simulations and thereby evaluate the variability of these models. While our
results are prior to calibration by the Phillips relation, understanding
these first-order effects on the brightness is a significant
step toward assessing the intrinsic scatter of these events. 

We also investigated explosions from hybrid C-O-Ne progenitors and
found that these progenitors produce explosions similar to
explosions from traditional C-O progenitors, but with a slightly lower
yield of \Ni and ejecta with slightly lower kinetic energy. The viability of 
these models is significant in that
the hybrid progenitors require less accreted mass and thereby
ameliorate a concern with the single-degenerate paradigm.

In the future, we will use these tools to explore simulations from
more realistic progenitors as these become available. We are  presently
performing a study of the convective Urca process with fully three-dimensional
simulations.
Introduced by Paczy{\'n}ski~\cite{paczynski72}, the convective Urca process
occurs when neutrino losses alter the dynamics of convection, in this case in
the simmering progenitor WD ~\cite{Iben78a,Iben78b,Iben82,Stein2006The-Convective-}. 
The process occurs in a region of the star known as the 
``Urca shell."  Above the density of the shell,
electron capture is favorable due to the high Fermi energy, while
below this density $\beta$-decay occurs.  Generally an Urca shell is specific to
an Urca pair of nuclides, and a pair that is important for thermonuclear
supernova progenitors is $^{23}$Ne/$^{23}$Na. The effect of the Urca shell
is twofold---energy is lost to neutrinos and the shell forms a barrier to 
buoyancy because of the different nucleon/electron ratio across the shell.
Through these two effects the Urca shell can influence both the energy content of the star and
the flow structure in the convection zone.
The results of our
study will inform progenitor models for future explosion simulations
that will address the effects of these results on the brightness of
events.

\ack

The work described in this paper is a distillation of 
parts of a large effort exploring thermonuclear supernovae.
The authors gratefully acknowledge contributions 
from Frank Timmes, Ed Brown, David Chamulak,
Daan van Rossum,
Shimon Asida, Ivo Seitenzahl, Fang Peng, Natalia Vladimirova,
Don Lamb, and Jim Truran. 
This work was supported in part by the US Department of Energy under
grants DE-FG02-07ER41516 and DE-FG02-87ER40317.
This work was supported in part by the US National Science 
Foundation under grant AST-0507456. 
This work was supported in part by the US National Aeronautics 
and Space Administration under grant NNX09AD19G. 
D.M.T. received support from the Bart J. Bok fellowship at
the University of Arizona for part of this work. 
A.P.J. received support from a National Research Council Research
Associateship for part of this work.
The authors acknowledge
the hospitality of the Kavli Institute for Theoretical
Physics, which is supported by the NSF under grant PHY05-
51164, during the programs ``Accretion and Explosion: the Astrophysics
of Degenerate Stars" and ``Stellar Death and Supernovae."
The software used in this work was in part
developed by the DOE-supported ASC/Alliances Center for Astrophysical
Thermonuclear Flashes at the University of Chicago. Results in this
paper were obtained using the high-performance computing system at the
Institute for Advanced Computational Science at Stony Brook
University and via US National Science Foundation TeraGrid and 
US Department of Energy INCITE awards. The authors also thank
Rachel Losacco and Ishmam Yousuf for previewing the manuscript.

\end{document}